\documentstyle[epsfig]{mn} 
\newcommand{\be}{\begin{equation}} 
\newcommand{\ee}{\end{equation}} 
\newcommand{\ba}{\begin{eqnarray}} 
\newcommand{\ea}{\end{eqnarray}} 
\newcommand{\r}{\mbox{\boldmath $r$}} 
\newcommand{\rhat}{\mbox{\boldmath $\hat{r}$}} 
\newcommand{\s}{\mbox{\boldmath $s$}} 
\newcommand{\vb}{\mbox{\boldmath $v$}} 
\newcommand{\q}{\mbox{\boldmath $q$}} 
\newcommand{\0}{\mbox{\boldmath $0$}} 
\newcommand{\xib}{\mbox{\boldmath $\xi$}} 
\newcommand{\kms}{{\rm km}s^{-1}} 
\newcommand{\mpc}{\,h^{-1}\!{\rm Mpc}} 
 
\title[PIZA and redshift surveys]{Reconstructing PSCz with a Generalised PIZA} 
\author[H.E.M. Valentine et al.] 
{Helen Valentine\thanks{hemv@roe.ac.uk}, Will Saunders and Andy Taylor\\ 
Institute for Astronomy, University of Edinburgh, Royal Observatory, Blackford Hill,Edinburgh, EH9~3HJ\\ }  

\begin{document}  
\maketitle 
\renewcommand{\baselinestretch}{1.0}
\begin{abstract} 
We present a generalized version of the Path Interchange Zel'dovich Approximation (PIZA), for use with realistic galaxy redshift surveys.  PIZA is a linear particle--based Lagrangian method which uses the present day positions of galaxies to reconstruct both the initial density field and the present day peculiar velocity field.    
We generalize the method by mapping galaxy positions from redshift--space to real--space in the Local Group frame, and take the selection function into account by minimizing the mass weighted action.   
We have applied our new method to mock galaxy catalogues, and find it offers an improvement in the accuracy of reconstructions over linear theory.  We have applied our method to the all-sky {\it IRAS} Point Source Catalogue Redshift Survey.   
Applying PIZA to PSCz, we are able to obtain the peculiar velocity field, the dipole, bulk velocities, and the distortion parameter.  From the dipole, we find that $\beta=0.51\pm0.14$.
We compare the PSCz bulk velocity with that of Mk III and conclude that $\beta=0.5\pm0.15$.  We  compare the PSCz dipole with that of SFI and find that $\beta=0.55\pm0.1$.
\end{abstract} 
\section{Introduction}  
Cosmological density and velocity fields may be reconstructed using the radial velocity information from a galaxy redshift survey.  To do this, assumptions must be made about the way the peculiar velocity field is produced, the relations between the velocity and density fields, and the relationship between galaxies and the underlying mass distribution.  The reconstructed fields may then be compared with the directly measured fields, e.g. from a peculiar velocity survey, and used to place constraints on cosmological parameters. 
 
\par The structures seen in the Universe today formed by the growth via gravitational instability of small density perturbations present in the early Universe.  The growth and evolution of these structures may be studied using methods of two types: Eulerian methods that solve the gravitational instability equations (see Peebles 1980); and Lagrangian methods that follow individual galaxy displacements.  These methods may be used to deduce the velocity field from the density field and vice versa. 
\par Eulerian methods include those which use linear perturbation theory applied to the gravitational instability equations.  The iterative reconstruction methods of Yahil et al. \shortcite{y}, who studied the 1.2Jy redshift survey, and Kaiser et al. \shortcite{k} and Taylor and Rowan-Robinson \shortcite{trr} who studied the QDOT redshift survey, use this method to reconstruct the peculiar velocity field from galaxy redshift surveys.  Branchini et al. \shortcite{b} use a similar method to reconstruct the peculiar velocity of the PSCz. 
\par Lagrangian methods include the Zel'dovich approximation (ZA; Zel'dovich 1970), which was suggested as a way to study the evolution of structure in the nonlinear regime.  Within the ZA the linear displacement field of galaxies is extrapolated from initial, Lagrangian coordinates to final, Eulerian coordinates. The galaxy peculiar velocity is simply the time derivative of the linear displacement. 

\par The use of Hamilton's principle, that the action is minimized during the evolution of a collection of particles under gravity, was introduced as a reconstruction method to cosmology by Peebles \shortcite{peeb2} and developed by Peebles (1994, 1995),  Shaya, Peebles and Tully \shortcite{spt}, Giavalisco et al. \shortcite{gmmy} and Susperregi and Binney \shortcite{sb}.  Sharpe et al. \shortcite{srr} recently used Peebles' numerical action method to reconstruct the peculiar velocities of PSCz galaxies within $cz=2000\kms$, while Monaco and Efstathiou \shortcite{me} have developed ZTRACE, an iterative reconstruction method based on Lagrangian perturbation theory.
 
\par A method combining the ZA and the least action principle is the Path Interchange Zel'dovich Approximation (PIZA) of Croft and Gazta\~{n}aga \shortcite{cg} (hereafter CG97). They showed that using the ZA the least action solution is that which minimizes the total mean square particle displacement between initial and final galaxy positions. Hence, given a set of final galaxy positions and a random distribution of initial coordinates, the least action trajectories are those that minimize the total displacement. 
 
Previously PIZA has been used to reconstruct the initial density fields from simulated real space density fields (CG97; Narayanan and Croft 1999).  Here we wish to apply PIZA to a redshift survey in order to reconstruct the peculiar velocity and real space density fields.  
We apply our new PIZA method to the Point Source Catalogue Redshift survey (PSCz; Saunders at al 2000), a survey of 15,000 {\it IRAS} galaxies to a flux limit of 0.6 Jy at 60$\mu m$, covering 84 per cent of the sky.

\par The outline of the paper is as follows.  We review PIZA in Section~\ref{sec-rev}.  In Section~\ref{sec-gals} we describe the problems encountered when applying PIZA to redshift surveys, and the generalizations we have made to the method in order to overcome these problems.   
In Section~\ref{sec-test} we test PIZA using PSCz-like simulations, and compare the predictions for the fields with those given by linear theory.  In Section~\ref{sec-pscz} we apply our new method to the PSCz survey.  In Section~\ref{sec-disc} we discuss our results and present our conclusions.

\section{Path Interchange Zel'dovich Approximation} \label{sec-rev} Here we briefly review the original PIZA method.  For more details, we refer the reader to Croft \& Gazta\~{n}aga \shortcite{cg}. 
\par We define a galaxy position in terms of its initial position and displacement:  
\begin{equation} \r(\q,t)=\q+\xi(\q,t),\label{eq:za}\end{equation}  
where $\r$ is the Eulerian coordinate, $\q$ is the initial Lagrangian coordinate and $\xib$ is the galaxy displacement.   
 
CG97 (\S 2.1) showed that the action is proportional to the mass--weighted squared particle displacement,  
\begin{equation} S=\sum_{i} m_i \xi_i^2.\label{eq:sum}\end{equation}   
where $m_i$ is the mass of the $i^{th}$ particle. 
The least action principle then requires the minimization of this quantity.     
In CG97 it is assumed that the galaxies are of equal mass, simplifying the minimization of equation (\ref{eq:sum}). Assuming the final, Eulerian galaxy positions are known, the initial positions are drawn from a homogeneous distribution and assigned at random to the Eulerian galaxy positions. We shall refer to the initial objects as PIZA particles.  It is not necessary that there is only one particle per galaxy;  CG97 found that increasing the number of particles per galaxy improved the accuracy of the reconstruction.   
Applying PIZA is very simple.  Two galaxies are then picked at random and their PIZA particles interchanged.  If this decreases $S$ the interchange is kept, and if not the particles are swapped back.  Interchanges are attempted until the rate of change of $S$ with attempted swaps, the cooling rate, is very low.

\section{PIZA with realistic galaxy redshift surveys}  
\label{sec-gals}   
The PIZA method cannot be immediately applied to a realistic galaxy redshift survey. In the first instance the radial positions of galaxies are estimated from their redshift, which introduces distortions in galaxy positions due to the peculiar velocities of galaxies. This distortion will depend on the rest frame of the observer, which we shall assume is the Local Group frame. To apply PIZA we must first deal with this distortion. Secondly most galaxy redshift surveys are flux limited, so that the observed number of galaxies drops off with distance as only the brightest galaxies can be seen. This affects the number density of galaxies, which must now be regarded as sampling the total galaxy population. This means that the galaxies must be assigned effective masses which must be included in the PIZA minimization. Finally, the sky coverage of galaxies may not be complete and some way must be included in the reconstruction to account for regions of incompleteness. In this Section we shall discuss each of these effects in turn.

\subsection{PIZA in redshift space} 
\label{sec-zspace} 
The peculiar motions of galaxies, $\vb$, mean that in the CMB frame the radial coordinate in redshift space is a mixture of spatial position and projected velocity, 
\be 
	\s(\r,t) = \r + (\rhat.\vb(\r,t)) \rhat, 
\ee 
where $\s$ is the redshift position. In the ZA the velocity field of galaxies is a linear extrapolation, and related to the displacement field by 
\be 
	\vb = \dot{\xib} = f(\Omega) \xib, 
\ee 
where $f \equiv d \ln \delta / d \ln a \approx \Omega^{0.6}$ is the growth rate of linear perturbations \cite{peeb}. 
If galaxies are a locally biased representation of the density field then  
$\delta_{\rm gal} = F[\delta_{\rm mass}]\approx b \delta_{\rm mass}$ to first order. We can then define a galaxy displacement field which is related to the velocity field by 
\be 
	\vb = \beta \xib \label{eq:four}
\ee 
where  
\be 
	\beta = \frac{f(\Omega)}{b} 
\ee 
is the redshift distortion parameter. The redshift space displacement  
field is related to the real space displacement field by \cite{th}
\begin{equation}  \xi_i^s={\cal P}_{ij}\xi_j  , \label{eq:reddisp} \end{equation}  
where ${\cal P}_{ij}\equiv\delta_{ij}^K+\beta\hat{r}_i\hat{r}_j$ 
is the redshift space projection tensor, and $\delta_{ij}^K$ is the Kronecker tensor.   
Inverting equation (\ref{eq:reddisp}), gives \cite{tv}: 
\begin{equation}  \xi_i={\cal P}_{ij}^{-1}\xi_j^s, \label{eq:realdisp} \end{equation}  
where  
\begin{equation}  
{\cal P}_{ij}^{-1}\equiv\delta_{ij}^K-\frac{\beta}{1+\beta}\hat{r}_i\hat{r}_j, 
\end{equation}  
is the inverse redshift space projection tensor. 
Squaring this gives us the expression for the linear displacement field in terms of 
linear redshift displacements and $\beta$; 
\begin{equation} 
 \xi^2=\xi^{s^2}-(1-\frac{1}{(1+\beta)^2})\xi_r^{s^2}, 
\label{eq:xir} 
\end{equation} 
where $\xi_r^s=\xib.\rhat$ is the displacement vector in redshift space projected along the line of sight.   Thus we must assume an a priori value for $\beta$.

\subsection{PIZA with Local Group frame redshifts} 
\label{sec-lg} 
 
Redshift space distortions should also take into account the motion of the observer, in this case the Local Group motion.  
The transformation from real to redshift space in the Local Group rest frame  
is \cite{tv}
\begin{equation}  
	\xi_i={\cal P}_{ij}^{-1}\xi_j^{s,LG}+(\delta_{ij}^K-{\cal P}_{ij}^{-1})\xi_j(\0)), 
\end{equation} 
where the super- or sub-script LG denotes a variable measured in the Local Group frame, and $\xib(\0)$  
is the displacement of the Local Group. 
The real-space displacement field can now be expressed in terms of redshift displacements 
in the Local Group frame as 
\begin{equation}  
	\xi^2=\xi^{s^2}_{LG}-\xi_{r,LG}^{s^2}+\frac{(\beta\xi_r(\0)+\xi_{r,LG}^s)^2}{(1+\beta)^2} 
\label{eq:xire} 
\end{equation} 
To apply equation (\ref{eq:xire}) we require the displacement of the Local Group.  
We find this by placing a galaxy representing the Local Group at the origin and using its displacement as the Local Group displacement.

\subsection{Incomplete Sky Coverage}  
The incomplete sky coverage may cause problems with PIZA, as with other reconstruction methods, since in these regions we do not know the true galaxy distribution. A number of choices for filling in these regions present themselves. One could populate the incomplete regions randomly on the sky, assuming no information or leave the regions empty of both particles and galaxies.  These are equivalent except for particles adjacent to the mask.  Alternatively, one could interpolate across using correlations in the density field to fill in the gaps. In reconstructing PSCz we use the Fourier Interpolation Scheme of Saunders et al. \shortcite{s2} which allows optimal, nonlinear interpolation across these regions. 
\subsection{PIZA with a selection function}  
\label{sec-mass} 
In a flux limited survey, the number density of survey objects decreases with increasing redshift, as galaxies at greater distances must be brighter to be seen above the flux limit.   
This is quantified by the survey selection function, $\phi(r)$, the expected number of galaxies seen at distance $r$ above the flux limit in the absence of clustering.   
 
In order to account for this change in the number density with distance, we 
pick initial PIZA particles with the same selection function as the galaxies. 
To take account of those galaxies that fall below the flux limit we assign  
each galaxy a mass of $1/\phi(r)$ and each PIZA particle a mass of  
\be 
	m_i = 1/(\nu \phi(r_i)), 
\label{eq:partmass} 
\ee  
where $\nu$ is the mean number of particles assigned per galaxy.  In this work we use 10 particles per galaxy.  We use $\nu=10$ in order to reduce the shot noise in the reconstruction.
Equations \ref{eq:xire} and \ref{eq:partmass} are used in equation \ref{eq:sum} to minimize the displacement squared. 
 
With the inclusion of particle masses it is possible for S to decrease but light particles to acquire unrealistically large trajectories.   We have therefore added an additional constraint that the rms displacement squared in redshift space should also decrease. 
 
\subsection{Second Order Effects} 
The masses we have assigned to the galaxies are based on the selection function at their redshift distance. Because of peculiar velocities, these can be in error. We therefore developed a modified version of PIZA, MASSPIZA, whereby after PIZA is run, particles are randomly reassigned from galaxies whose mass has been overestimated to those where it has been underestimated. PIZA is then rerun and the whole process iterated until the masses are acceptable and the trajectories minimised. Unfortunately, this procedure is in general unstable to the rocket effects described by Kaiser (1987). In the section below, we show that PIZA without mass reassignment gives acceptable results in comparison with simulations; the agreement is in general not improved by the second, MASSPIZA stage. Therefore, we have not incorporated it in any of the results quoted in this paper.

\section{Tests on simulations}  
\label{sec-test} 
 
\begin{figure} 
\centerline{\epsfig{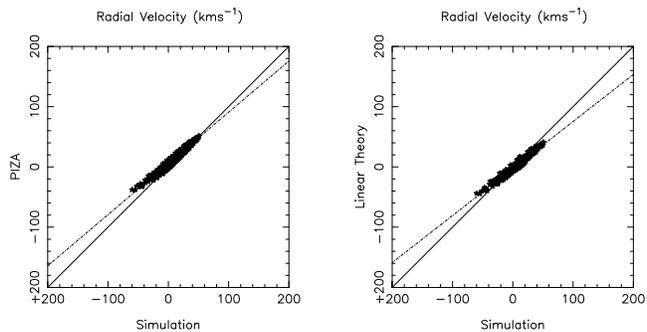}} 
\caption{Typical simulation results: comparison of PIZA and linear theory reconstructions. The left hand panel shows PIZA radial velocities compared with simulation, the right hand shows linear theory compared with simulation. Velocities were smoothed with a Gaussian of radius $20 \mpc$.  The dotted line indicates the least-squares fit. } 
\label{fig:shot1} 
\end{figure}

In order to test the new PIZA method, we have applied it to a set of PSCz-like mock catalogues.    
For more details of these mock catalogues, we refer the reader to Branchini et al. \shortcite{b2}.   
The catalogues are of two CDM cosmologies, a critical model , $\Omega_m=1$, $\beta=1$, with $\Gamma =0.25$  and a flat model with $\Omega_m =0.3$ and $\Omega_{\Lambda}=0.7$, $\beta=0.50$ and are limited to $v<15000 kms^{-1}$.  
We test the reconstruction of the radial peculiar velocity field, comparing it with that of the simulation. To compare the fields the peculiar velocities were binned in real space and smoothed with a Gaussian smoothing kernel.  We have also compared the PIZA reconstructions with the linear theory reconstructions for the mock catalogues from Branchini et al. 
 Figure~\ref{fig:shot1} shows typical results comparing PIZA and linear theory reconstructions with the simulations.  The dotted line indicates the the least squares fit to the data.   
As stated above, we must assume an a priori value for $\beta$.  We find that the reconstruction works well for any sensible value of $\beta$.
We find that generally the gradient reconstructed by PIZA is less biased than that reconstructed by linear theory but that the scatter is somewhat worse.  This is probably one of the limitations of PIZA as the scatter in redshift space reconstructions produced by PIZA is quite large (see also CG97). 
 
We find that PIZA works better for low density models as they are less dynamically evolved.

The accuracy of the reconstruction varies between different sets of initial conditions (i.e. PIZA particles) due to the random positioning of particles. There is also a scatter between reconstructions using the same initial conditions due to the random swaps made by PIZA. This probably reflects that there is not a simple minimum for the action, $S$, but that the configuration space is complicated by the particle discreteness. 

We estimate $\beta$ from the trajectory of the Local Group galaxy using equation \ref{eq:four} and assuming the magnitude of the CMB dipole.
Table \ref{tbl:betafracf} shows the reconstructed $\beta$, the dipole misalignment angle, and the slope of the peculiar velocity comparison, with the scatter due to different initial conditions.
We have tested the effects of having the Galactic Plane filled or empty.  This is important for PSCz as we know there are large structures, such as the Great Attractor, in the Plane.  With the simulations we have tested PIZA using $4\pi$ sky coverage, and with the PSCz mask applied.  With the mask empty we find that the average  misalignment angle is $24^{\circ}$, and with full-sky coverage it is $14^{\circ}$.  

\begin{table} 
\begin{tabular}{|l||l|l|l|l|l||l|}\hline 
Mock&$\beta_R$&Scatter&$\theta_{CMB}$&Slope&Scatter\\\hline\hline 
$\Lambda$0.7 P&0.42&0.07&22&0.76&0.10\\ \hline
$\Gamma$0.25 P&0.95&0.27&26&0.61&0.14\\ \hline
$\Lambda$0.7 $4\pi$&0.43&0.05&14&0.74&0.04\\ \hline 
$\Gamma$0.25 $4\pi$&0.84&0.18&13&0.65&0.13\\ \hline 
\end{tabular} 
\caption{Results from tests on simulations. $\Lambda$0.7 denotes the  $\Omega_{\Lambda}$ model, and $\Gamma$0.25 the critical model. P denotes simulations with PSCz sky coverage and $4\pi$ simulations with full sky coverage} 
\label{tbl:betafracf} 
\end{table} 
\section{Reconstructing PSCz}  
\label{sec-pscz} 
 
We have applied our new PIZA method to the PSCz survey (Saunders {\em et al.} 2000) out to $R=300 \mpc$, using the parametrized {\it IRAS} selection function of Mann, Saunders and Taylor \shortcite{mst}.  For the reconstructions we have filled the masked regions with interpolated galaxies.
 
\begin{figure} 
\centerline{\epsfig{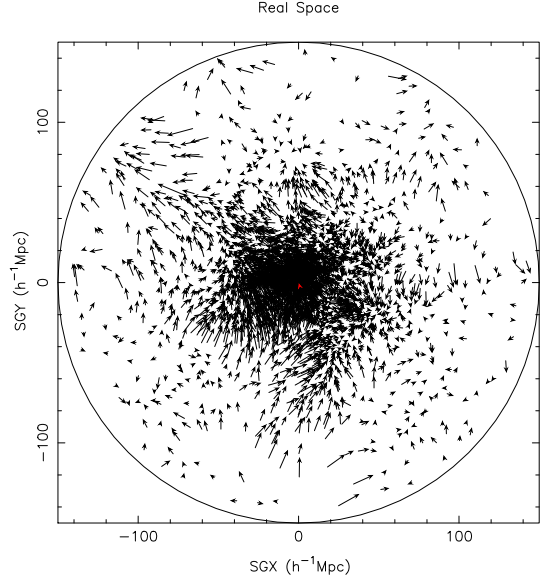}} 
\caption{PSCz velocity field within $150\mpc$} 
\label{fig:psczmaskvel} 
\end{figure}

\begin{figure} 
\centerline{\epsfig{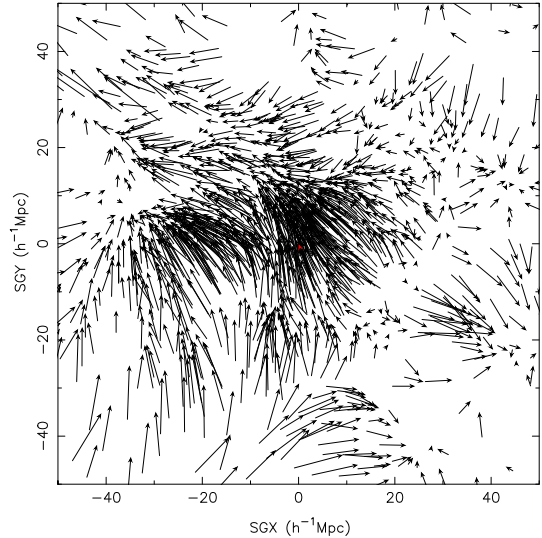}} 
\caption{PSCz velocity field within $50\mpc$} 
\label{fig:psczmaskvel2} 
\end{figure}

Figure~\ref{fig:psczmaskvel} shows the reconstructed PSCz velocity field in a slice $20 \mpc$ thick centred on the Supergalactic Plane.  Figure~\ref{fig:psczmaskvel2} shows the same slice out to $50\mpc$.  The arrows show the galaxy trajectories, with the arrows starting at the initial positions and the arrow heads at the galaxy real space positions. Generally the velocity field at the edge of the volume is quiet, apart from flow towards Shapley at (120,90) and the void at (0,-120).  The Coma cluster is seen at (0,70).  The Cetus wall extends northwards from (0,-100) towards Perseus-Pisces at (50,-20).  Virgo is seen at (0,10).  The Great Attractor is at (-40,20), and we see little evidence of backfall on to it on the opposite side to the G.A. from the Local Group.

\subsection{Cosmological dipole} 
We find an average dipole direction of (l,b)=(264.4$^{\circ}$,41.7$^{\circ}$).  We find that $\beta=0.51\pm0.14$.  This error has 3 sources:  random error due to different initial conditions;  misalignment between the reconstructed and CMB dipoles; cosmic variance and shot noise on the reconstructed dipole \cite{tv}.   A similar result is found by only using the z component of the dipole.

\begin{figure} 
\centerline{\epsfig{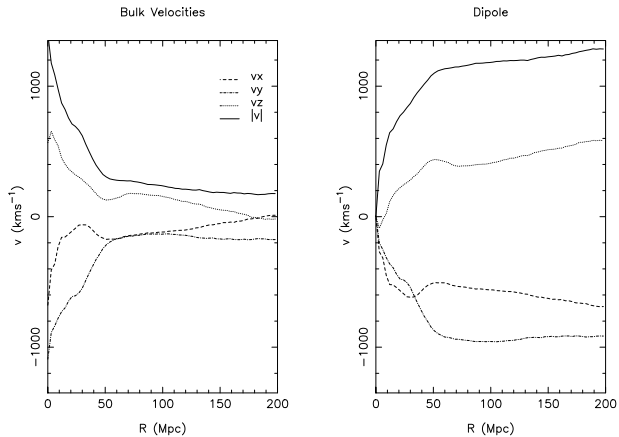}} 
\caption{PSCz bulk velocities and dipole within $150\mpc$} 
\label{fig:bulk} 
\end{figure} 
 
\subsection{Radial velocity field} 
The reconstructed dipole and bulk flows are shown in Figure~\ref{fig:bulk}.  We have compared our bulk flows with those from Mark III and the linear theory PSCz reconstruction (Branchini {\em et al.} 1999), and conclude that $\beta=0.5\pm0.15$. This is consistent with the results of Tadros et al (1999) based on the measurement of the degree of redshift space distortion in the PSCz. 
 
\subsection{Comparison with SFI}
We have undertaken a comparison of the reconstructed PSCz velocity field with the SFI data set of peculiar velocities of spiral galaxies (Giovanelli {\em et al.} 1998).  We have calculated the average peculiar velocity, with respect to the Local Group, of galaxies in redshift slices, to be as closely as possible comparable to Giovanelli {\em et al.}.  We find reasonable agreement in amplitude, though there are significant differences in direction, for $\beta=0.55\pm0.1$.  Table \ref{tbl:sfi} shows the average peculiar velocity in shells for PIZA and SFI.
\begin{table} 
\centering 
\begin{tabular}{|l|l|l|l|l|}\hline 
Shell&$V_{SFI}$&(l,b)$_{SFI}$&$V_{PIZA}$&(l,b)$_{PIZA}$\\ \hline\hline
   0 - 2000&270$\pm$80&(245,49)$\pm$19&372&(228,45)\\
1500 - 3500&410$\pm$69&(255,21)$\pm$12&430&(254,56)\\
2500 - 4500&620$\pm$76&(255,15)$\pm$11&485&(271,50)\\
3500 - 5500&585$\pm$92&(265,19)$\pm$13&550&(281,39)\\
4500 - 6500&544$\pm$98&(270,16)$\pm$15&503&(279,34)\\
\end{tabular}  
\caption{SFI vs PSCz peculiar velocities with respect to shells of galaxies.  PSCz reconstruction assumes $\beta=0.5$.} 
\label{tbl:sfi} 
\end{table} 
\section{Discussion}
\label{sec-disc}   
We have presented a generalized version of the PIZA method for reconstructing velocity and density fields from galaxy surveys.  These modifications to the original method are necessary if it is to be applied to realistic galaxy redshift surveys with a selection function.  We have used the inverse redshift space operator of Taylor and Valentine \shortcite{tv} to write the mean square particle displacement in terms of the redshift space positions of galaxies, and to map redshifts from the Local Group rest frame to real space. The PIZA particles obey the same selection function as the galaxies. Doing this we can reconstruct the real space density field, the galaxy initial positions and the peculiar velocity field.   
 
We have tested the method using PSCz-like CDM simulations, testing the reconstruction of the real space positions and the peculiar velocities.  We have also compared our reconstructions with linear theory reconstructions of the simulations. 
We have applied our method to the PSCz survey, and reconstructed the local velocity field out to $300 \mpc$ from the Local Group, and have calculated the bulk flow and the dipole.  We find the reconstructed dipole to be $16^{\circ}\pm4$ away from the CMB dipole.  From comparison between the PIZA reconstruction and the Mark III bulk flow, we find that $\beta=0.5\pm0.15$, while from a dipole analysis of the PIZA reconstruction, we find that $\beta=0.51\pm0.14$.

\section{Acknowledgements} 
HEMV thanks PPARC for a studentship.  We thank Enzo Branchini for the simulations and linear theory reconstructions.

\end{document}